\documentclass[a4paper,11pt]{article}
\usepackage{pos}
\setcitestyle{maxnames=1}

\usepackage{subcaption, graphicx}
\usepackage{graphicx}
\usepackage{bbm}
\newcommand{\oo}{\ensuremath{\mathrm{O}}}

\title{Nucleon-sigma terms at $m_\pi=222 ~\rm MeV$ with a variational analysis from lattice QCD}

\author*[a]{Lorenzo Barca}
\author[b]{Gunnar Bali}
\author[b]{Sara Collins}
\author[b]{Marcel Rodekamp}

\affiliation[a]{John von Neumann-Institut für Computing NIC, Deutsches Elektronen-Synchrotron DESY, \\ Platanenallee 6, 15738 Zeuthen, Germany}

\affiliation[b]{Fakultät für Physik, Universität Regensburg, Universitätsstr. 31, 93053 Regensburg, Germany}

\emailAdd{lorenzo.barca@desy.de}
\emailAdd{gunnar.bali@ur.de}
\emailAdd{sara.collins@ur.de}
\emailAdd{marcel.rodekamp@ur.de}

\abstract{
Nucleon sigma terms are important for the decomposition of the nucleon mass and for searches for new physics 
beyond the Standard Model involving scalar interactions. A persistent tension between lattice QCD and 
phenomenological determinations may be due to uncontrolled excited-state contamination in lattice QCD analyses. 
In previous work at $m_\pi=429~\rm MeV$, we showed that this contamination is dominated by $N\sigma$ states and 
can be strongly suppressed with a variational analysis using $N$ and $N\sigma$ operators. In this talk, we present 
preliminary results at $m_\pi=222~\rm MeV$, where the $\sigma$ becomes unstable and decays into $\pi\pi$. 
We investigate whether the same small variational basis can remove the expected $N\pi\pi$ contamination 
through the overlap of the $N\sigma$ operator with these states.
}

\FullConference{The 43rd International Symposium on Lattice Field Theory (LATTICE2026)\\
26 July - 1 August 2026\\
University of Maryland, USA \\}

\begin{document}
\begin{flushright}
    \texttt{DESY-26-116}
\end{flushright}
\maketitle

\section{Introduction}
The nucleon sigma terms quantify the contribution of explicit chiral
symmetry breaking to the nucleon mass. For a quark flavour q, they
are defined as
\begin{equation}
\sigma_{qN}
=
m_q \langle N | \bar q q | N \rangle
=
m_q \frac{\partial m_N}{\partial m_q},
\label{eq:sigma_definition}
\end{equation}
where the second equality follows from the Feynman--Hellmann theorem.
In the isospin-symmetric limit, $m_u=m_d=\hat m$, the pion--nucleon
sigma term is $\sigma_{\pi N} =\hat m \langle N | \bar u u+\bar d d | N\rangle$.
The sigma terms determine the effective coupling of scalar particles,
including the Standard Model Higgs boson, to the nucleon at zero
momentum transfer. They are therefore important hadronic inputs for
Standard Model phenomenology and searches for physics beyond it,
including dark-matter--nucleon scattering through scalar interactions
\cite{Alarcon:2021dlz}.

Phenomenological determinations of $\sigma_{\pi N}$ are obtained from
pion--nucleon scattering using low-energy theorems, dispersion
relations and chiral perturbation theory. A high-precision analysis
based on the Cheng-Dashen low-energy theorem, pionic-atom data and
Roy--Steiner equations yields $\sigma_{\pi N}=59.1(3.5)~\mathrm{MeV}$
\cite{Hoferichter:2015dsa}. Lattice QCD provides a
first-principles determination, either directly from scalar nucleon
matrix elements or indirectly from the quark-mass dependence of the
nucleon mass through Eq.~\eqref{eq:sigma_definition}. A tension remains
between phenomenology and most lattice results: the FLAG
$N_f=2+1$ average is $\sigma_{\pi N}=42.2(2.4)~\mathrm{MeV}$, whereas the
$N_f=2+1+1$ average, $\sigma_{\pi N}=60.9(6.5)~\mathrm{MeV}$, 
is compatible with phenomenology but is dominated by the PNDME determination
\cite{FLAG2024,Gupta:2021ahb}.

Differences in the isospin conventions employed in phenomenological
and lattice calculations account for a correction
$\Delta\sigma_{\pi N}=3.1(5)~\mathrm{MeV}$ \cite{Hoferichter:2023ptl}, 
which reduces but does not resolve the discrepancy. 
The treatment of the quark-mass dependence may provide an
additional source of uncertainty. A recent
$\mathrm{SU}(2)$ baryon chiral perturbation theory analysis at
$\mathcal{O}(p^5)$, including isoscalar $\pi\pi$ rescattering,
obtained $\sigma_{\pi N}=55.9(2.5)~\mathrm{MeV}$ from existing
$N_f=2+1$ lattice data, compared with
$48.1(1.9)~\mathrm{MeV}$ at leading one-loop order
\cite{Liang:2025adz}. This suggests that scalar-isoscalar two-pion
dynamics may be important in the chiral extrapolation.

A controlled lattice determination also requires continuum and
finite-volume extrapolations and a reliable separation of the nucleon
ground state from excited-state contributions (ESC). The latter is
particularly challenging in direct calculations. Guided by chiral
perturbation theory, the PNDME collaboration constrained the lowest
energy gaps in its spectral analysis using non-interacting $N\pi$ and
$N\pi\pi$ energies \cite{Gupta:2021ahb}. Near the physical pion mass,
including these states changed their result from
$\sigma_{\pi N}=41.9(4.9)~\mathrm{MeV}$) in a conventional analysis
to $59.6(7.4)~\mathrm{MeV}$. An insufficient treatment of
multi-hadron excited states may therefore contribute to the tension
between lattice and phenomenological determinations.

As discussed in Ref.~\cite{Barca:2025det}, excited states that can be
created directly by the inserted current need not exhibit the usual
finite-volume suppression expected for multi-hadron contamination in
three-point functions. For an isoscalar scalar current, the relevant
states contain a nucleon together with a scalar-isoscalar excitation.
At heavy pion masses this may be described as an $N\sigma$ state,
whereas near the physical point the channel is dominated by interacting
two-pion states associated with the broad
$\sigma/f_0(500)$ resonance.

In Ref.~\cite{Barca:2024hrl}, we studied this mechanism using a
variational basis of conventional nucleon and $N\sigma$ interpolating
operators. On an $N_f=3$ ensemble with
$m_\pi=429~\mathrm{MeV}$, where the lowest scalar meson is stable,
the dominant ESC in isoscalar scalar nucleon three-point functions was
identified with the transition between the nucleon and an
S-wave $N(0)\sigma(0)$ state. Including $N\sigma$-type
interpolators in a generalized-eigenvalue analysis substantially
reduced this contamination and allowed the scalar charges to be
extracted at shorter source--sink separations.

In these proceedings, we extend this study to
$m_\pi=222~\mathrm{MeV}$, where the scalar-isoscalar channel is
dominated by two-pion dynamics rather than a stable sigma meson. We
investigate whether a variational basis containing nucleon and
$N\sigma$ interpolators can resolve and suppress the leading
$N\pi\pi$ contamination in nucleon scalar three-point functions. We
also discuss the general mechanism responsible for current-enhanced
states and present preliminary results for the light- and strange-quark
scalar matrix elements.
 \section{Current-enhanced states in lattice calculations of nucleon matrix elements}

Scalar matrix elements are extracted from Euclidean three-point
correlation functions,
\begin{equation}
C^{S^q}_{\mathrm{3pt}}(t,\tau)
=
\langle \oo_N(t) S^q(\tau) \bar{\oo}_N(0)\rangle
-
\langle \oo_N(t)\bar{\oo}_N(0)\rangle
\langle S^q(\tau)\rangle ,
\qquad
S^q(\tau)=\bar q 1 q,
\label{eq:c3pt}
\end{equation}
where all the operators are projected to zero-momentum and the second 
term subtracts the vacuum expectation value associated with the scalar current. 
Defining $Z_n=\langle 0|\oo_N|n\rangle$, its spectral decomposition is
\begin{align}
C^{S^q}_{\mathrm{3pt}}(t,\tau)
=&~
|Z_N|^2
\langle N|S^q|N\rangle e^{-m_Nt}
+
\sum_{n\neq N}
Z_N\bar Z_n
\langle N|S^q|n\rangle
e^{-m_N(t-\tau)}e^{-E_n\tau}
+
\\
\nonumber\
+&
\sum_{n'\neq N}
Z_{n'}\bar Z_N
\langle n'|S^q|N\rangle
e^{-E_{n'}(t-\tau)}e^{-m_N\tau}
+
\sum_{n',n\neq N}
Z_{n'}\bar Z_n
\langle n'|S^q|n\rangle
e^{-E_{n'}(t-\tau)}e^{-E_n\tau}.
\label{eq:spectral_decomposition}
\end{align}
Finite-volume normalization factors are absorbed into
the overlaps and matrix elements.
Using the nucleon two-point function
$C_{\mathrm{2pt}}(t)
=
\langle \oo_N(t)\bar{\oo}_N(0)\rangle$,
we construct the standard ratios
\begin{equation}
\label{std_ratios}
R^{S^q}(t,\tau)
=
\frac{C^{S^q}_{\mathrm{3pt}}(t,\tau)}
{C_{\mathrm{2pt}}(t)}
=
g_S^q
~+
\mathcal{O}(e^{-\Delta E_{n'}(t-\tau)}e^{-\Delta E_n\tau})
~
(\rm ESC)
\end{equation}
where $g_S^q$ is the scalar nucleon quark charge which parametrises $\langle N|S^q|N\rangle$,
$\Delta E_n=E_n-m_N$, and $\Delta E_{n'}=E_{n'} - m_N$. 
Ground-state dominance is formally reached for large source--current and 
current--sink separations. In practice, the
exponentially deteriorating signal-to-noise (StN) ratio of baryonic
correlation functions limits the accessible Euclidean times, and ESC
may remain significant. Multilevel sampling algorithms have the potential to mitigate exponentially
the StN \cite{Barca:2025dca}.

Following the argument of Refs.~\cite{Barca:2025det, Barca:2026juc}, we expect sizeable
ESC from multi-hadron states $|NM\rangle$, where M carries
scalar-isoscalar quantum numbers and the same momentum as the current. 
At heavy pion masses, the relevant state is an $|N\sigma\rangle$ state, while at lighter pion masses the
scalar-isoscalar channel is expected to be dominated by $|N\pi\pi\rangle$ scattering states
with $\pi\pi$ in S-wave.
 \section{Variational analysis with $N$ and $N\sigma$-like interpolating operators}
\begin{figure}[t]
\centering
\begin{subfigure}[t]{0.49\linewidth}
  \centering
  \includegraphics[width=\linewidth]{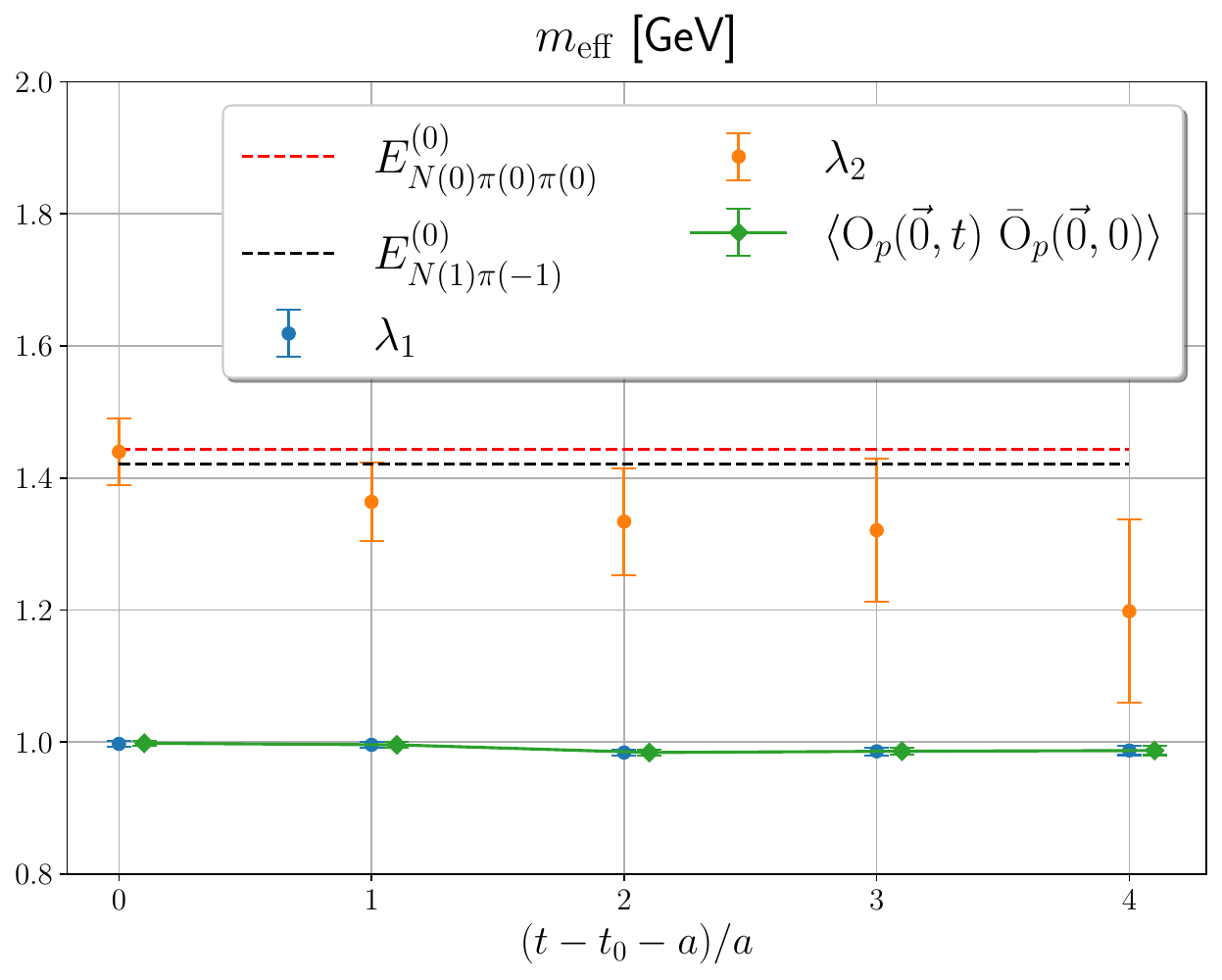}
\end{subfigure}
\hfill
\begin{subfigure}[t]{0.49\linewidth}
  \centering
  \includegraphics[width=\linewidth]{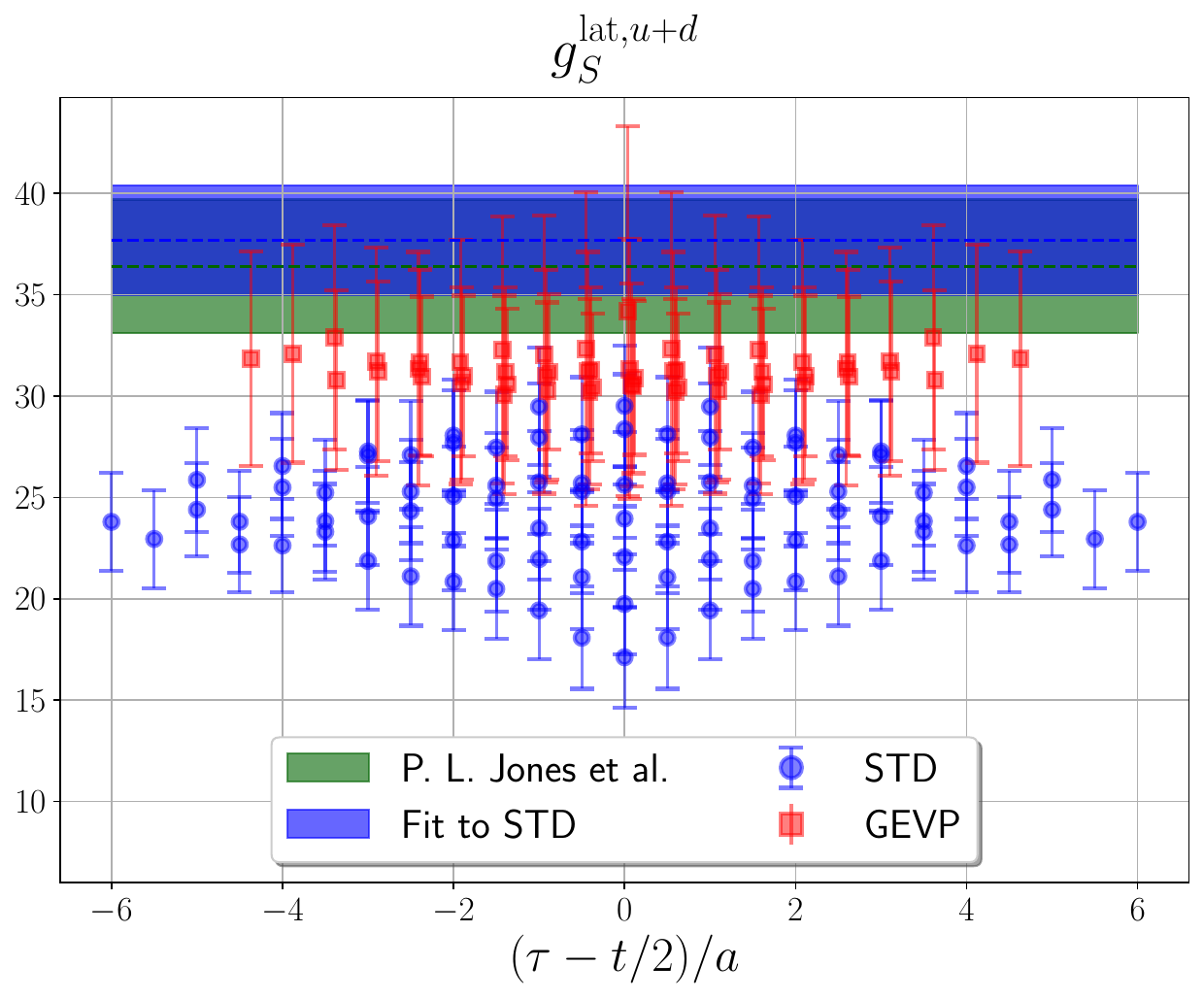}
\end{subfigure}
\caption{
(left) Effective masses of the GEVP eigenvalues $\lambda_1$ and
$\lambda_2$, obtained with the reference time
$t_0=7a\approx0.6~\rm fm$, compared with the nucleon effective mass
and the lowest non-interacting $N\pi$ and $N\pi\pi$ energies.
(right) Comparison of the standard (blue, STD) and GEVP-improved
(red, GEVP) ratios at different source-sink separations, starting from
$t=4a\approx0.34~\rm fm$. Two time slices adjacent to both the source
and the sink are omitted. The blue band denotes the result of a fit to
the summed standard ratios in Eq.~\eqref{std_ratios}.
}
\label{fig:gevp}
\end{figure}

Motivated by this expectation, we perform a variational analysis using $N$- and $N\sigma$-type interpolating operators on two CLS ensembles at $m_\pi=429~\rm MeV$ (A653) and $222~\rm MeV$ (C101), see Tab~\ref{tab:ensembles} and \cite{RQCD:2022xux}.
The interpolating operators that are used are the standard 3-quark nucleon interpolator for $\oo_N$ and $\oo_{N\sigma}=\oo_N(\vec{0}) \oo_\sigma(\vec{0})$ with $\oo_\sigma= \left(u\bar{u} + d\bar{d}\right) / \sqrt{2}$.
The nucleon interpolators $\oo_N$ are constructed using
Wuppertal-smeared quark fields and APE-smeared gauge links, resulting
in a smearing radius $\sqrt{\langle r^2\rangle}\approx1.0~\rm fm$. The scalar
interpolator $\oo_\sigma$ is smeared with a smaller radius,
$\sqrt{\langle r^2\rangle}\approx0.20~\rm fm$. These choices are
designed to enhance the overlap with the low-lying nucleon and scalar
states, respectively.
Using the operator basis $\mathbb{B}=\left\{ \oo_N, \oo_{N\sigma} \right\}$, we construct a matrix of two-point correlation functions $C_{\rm 2pt}(t)_{ij} = \langle \oo_i(t) ~\bar{\oo}_j(0)\rangle$ with $\oo_i, \oo_j \in \mathbb{B}$, and solve the generalised eigenvalue problem (GEVP) at different reference times $t_0$
\begin{equation}
\label{gevp}
C_{\rm 2pt}(t) V(t,t_0)=C_{\rm 2pt}(t_0) V(t,t_0) \Lambda(t,t_0),
\qquad t>t_0.
\end{equation}
This gives the matrix of generalised eigenvalues $\Lambda(t,t_0)=\mathrm{diag}(\lambda^\alpha(t,t_0))$
and eigenvectors $V(t,t_0)=(v_{\oo_i}^\alpha(t,t_0))$, where the superscripts (subscripts) refer to the eigenstate (operator).
In the limit of large reference times $t_0$ and $t$, the eigenvalues decay exponentially with the energy of the state, $\lambda^\alpha(t,t_0)\propto e^{-E_\alpha(t-t_0)}$, while the components of the eigenvectors $v_i^\alpha(t,t_0)$ are related to the overlap of the operator $\oo_i$ with the state $\alpha$.
An operator with an improved overlap to the ground state can be constructed from a linear combination of the initial operators $\oo_i$ and the eigenvectors for the first level $v_i^{\alpha=1}(t', t_0)$ evaluated at a sufficiently large time $t'$ and $t_0$:
\begin{equation}
\label{gevp_op}
\oo^{\rm imp}_N(t) = \sum_{\oo_i \in \mathbb{B}} v_{\oo_i}^{\alpha=1}(t',t_0)~\oo_i(t).
\end{equation}
This improved operator can be used to define GEVP-improved two-point functions $C^{\rm imp}_{\rm 2pt}(t)$ and three-point functions $C^{\rm imp, S^q}_{\rm 3pt}(t,\tau)$,
and to construct GEVP-improved ratios
\begin{equation}
\label{gevp_ratios}
R^{\rm imp, S^q}(t,\tau)
=
\frac{C_{\rm 3pt}^{\rm imp, S^q}(t,\tau)}{C_{\rm 2pt}^{\rm imp}(t)}.
\end{equation}
\begin{table}[h]
\centering
\begin{tabular}{ccccccc}
\hline
Ensemble & \(m_\pi\,[\mathrm{MeV}]\) & \(m_K\,[\mathrm{MeV}]\) & \(a\,[\mathrm{fm}]\)
         & \(L/a\) & \(T/a\) & \(N_{\rm cfg}\) \\
\hline
A653 & 429 & 429 & 0.098  & 24 & 48 & 800  \\
C101 & 222 & 476 & 0.0865 & 48 & 96 & 2000 \\
\hline
\end{tabular}
\caption{Parameters of the CLS ensembles used in this work.}
\label{tab:ensembles}
\end{table}
\paragraph{Variational analysis on ensemble A653 \cite{Barca:2024hrl}}-- A preliminary analysis of the zero-momentum
scalar isoscalar two-point function yields an energy $m_\sigma=550(50)~\rm MeV$, 
well below the lowest non-interacting $\pi\pi$ energy. 
The lowest scalar state can therefore be regarded as a stable $\sigma$ meson on this ensemble.
The GEVP solutions of Eq.~\eqref{gevp} yield a lowest level consistent
with the nucleon mass and a second level consistent with the
non-interacting $N(0)\sigma(0)$ energy; see the left panel of Fig.~1
in Ref.~\cite{Barca:2024hrl}. On this ensemble, the lowest $N\pi$
and non-interacting $N\pi\pi$ energies lie above these two levels
\cite{Barca:2022uhi}. Their contributions to the truncated GEVP are
therefore expected to become increasingly suppressed at sufficiently
large Euclidean times.
Using the GEVP eigenvectors, we construct the improved two- and
three-point functions of Eq.~\eqref{gevp_op}. In the projection of the
three-point correlation matrix, we neglect the
$N\sigma\to N\sigma$ term $\langle
\oo_{N\sigma}(t)S^q(\tau)\bar{\oo}_{N\sigma}(0)
\rangle$, which is expected to be subleading compared with the
$N\leftrightarrow N\sigma$ transition contributions.
As shown in Fig.~2 of Ref.~\cite{Barca:2024hrl}, this two-dimensional
variational basis substantially suppresses the ESC. The GEVP-improved
ratios are consistent with a plateau within statistical uncertainties
already for $t\gtrsim0.5~\rm fm$ and $0.2~\rm fm<\tau<t-0.2~\rm fm$.
Notice that the same ensemble was used to analyse the dominant $N\pi$ ESC 
in nucleon axial and pseudoscalar three-point functions via a variational 
analysis with $N$ and $N\pi$-type interpolators \cite{Barca:2022uhi},
and to identify the dominant $N\rho$ contribution in nucleon vector isovector 
three-point functions via a GEVP.
\paragraph{Variational analysis on C101}-- At $m_\pi=222~\rm MeV$, 
the zero-momentum scalar-isoscalar two-point function decays with an energy of approximately
$410~\rm MeV$, close to the lowest non-interacting $\pi\pi$
energy. At this lighter pion mass, on the CLS trajectory of constant average
sea-quark mass,
\(\bar m=(2m_\ell+m_s)/3=\mathrm{const.}\) (see Tab.~\ref{tab:ensembles}), 
we expect the $\sigma$ to be unstable and the $\pi\pi$ decay channel to be open.
\footnote{At a similar pion mass,
$m_\pi=236~\mathrm{MeV}$, the HadSpec Collaboration finds the $\sigma$
to be a broad resonance~\cite{Briceno:2016mjc}.}
Nevertheless, we construct a basis of $N$- and $N\sigma$-like
interpolating operators and solve the GEVP in Eq.~\eqref{gevp} at
different reference times. The left panel of Fig.~\ref{fig:gevp}
compares the effective masses of the GEVP eigenvalues
$\lambda_1$ and $\lambda_2$, obtained at $t_0=7a\approx0.6~\rm fm$, 
with the nucleon effective mass and the lowest non-interacting 
$N\pi$ and $N\pi\pi$ energies. The first eigenvalue is consistent 
with the nucleon ground state, whereas the second lies close to 
the nearly degenerate $N\pi$ and $N\pi\pi$ levels.

To construct the improved interpolator, we use the eigenvectors at
$t_0=7a$, averaged over $t-t_0-a=5a,6a,7a$, and form the GEVP-improved 
ratios defined in Eq.~\eqref{gevp_ratios}. The right panel of Fig.~\ref{fig:gevp}
compares the standard and GEVP-improved ratios at source-sink
separations starting from $t=4a\approx0.34~\rm fm$. Including the
$N\sigma$-like interpolator significantly suppresses the ESC, and
the improved ratios approach the reference determinations at
considerably shorter source-sink separations. The blue band shows our
fit to the summed standard ratios $S(t) = \sum_{\tau=2a}^{t-2a} R(t,\tau)$ in Eq.~\eqref{std_ratios}, 
which yields the unrenormalised scalar charge $g_S=37.7(2.7)$.
In the model average we use $N\pi$ and $N\pi\pi$ fit priors as well as 
the energy gaps from the nucleon two-point functions to constraint the first excited state, 
as discussed in Ref.~\cite{Barca:2024hrl}. A simple linear fit to the summed GEVP ratios gives
$g_S=31.2(4.0)$, consistent with the fit to the standard ratios within $1.6\sigma$.
The green band shows the result of an independent standard analysis performed 
on the same ensemble \cite{Pia:lattice2026}.

 \section{Conclusions}
We have investigated nucleon sigma terms using a variational basis of
$N$- and $N\sigma$-like interpolating operators at $m_\pi=429~\rm MeV$ and $222~\rm MeV$.
The results at $m_\pi=429~\rm MeV$, first reported in
Ref.~\cite{Barca:2024hrl}, correspond to a regime in which the lowest
scalar state is stable, with $m_\sigma=550(50)~\rm MeV$. 
The $N(0)\sigma(0)$ state is then both the current-enhanced and the
lowest relevant multi-hadron excitation and lies below the lowest
non-interacting $N\pi$ and $N\pi\pi$ energies. In this case, the
$2\times2$ GEVP basis substantially suppresses the ESC already at
source-sink separations $t\gtrsim0.5~\rm fm$. The resulting
GEVP-improved ratios are constant within statistical uncertainties for
insertion times $0.2~\rm fm<\tau<t-0.2~\rm fm$.
The corresponding analysis at $m_\pi=222~\rm MeV$ is presented here
for the first time. At this lighter pion mass, the scalar-isoscalar
channel is governed by two-pion dynamics rather than by a stable
$\sigma$ meson, and the relevant current-enhanced excitations are
expected to be $N\pi\pi$ finite-volume states. The
$N\sigma$-like interpolator nevertheless remains effective because
it overlaps with these states. The GEVP-improved ratios display
visibly reduced ESC relative to the standard ratios already at
$t\approx0.4~\rm fm$ and approach the long-distance determinations
at shorter source-sink separations. Their statistical uncertainties
are currently larger, however, and are dominated by the noise of the
GEVP eigenvectors entering the optimized linear combination.
These results provide evidence that including interpolators designed
to couple to the current-enhanced states identified in
Ref.~\cite{Barca:2025det} is an effective strategy for suppressing ESC.
Importantly, the method remains successful at lighter pion masses,
where the relevant scalar-isoscalar excitation belongs to a resonant
two-pion channel rather than corresponding to a stable particle. This
also provides a natural interpretation of the results of
Ref.~\cite{Alexandrou:2024tin}, where a GEVP basis containing $N$- and
the lowest $N\pi$-like interpolators did not significantly improve the
suppression of ESC in nucleon sigma-term calculations even at physical pion masses. 
The scalar-isoscalar current cannot create a single pion directly from the
vacuum, and the $N\pi$ states are therefore not enhanced by the
same mechanism. However, the contribution of the tower of $N\pi$ states
can be non-negligible.
Including $N\pi$-like interpolators remains important
for obtaining a more complete description of the low-lying nucleon
spectrum. On the C101 ensemble, the lowest $N\pi$ and $N\pi\pi$
levels are nearly degenerate. Omitting the nearby
$N(-\vec p)\pi(\vec p)$ state may therefore introduce systematic
uncertainties in the extracted eigenvectors. Extending the basis to a
$3\times3$ GEVP containing $N$-, $N\sigma$- and $N\pi$-like
interpolators should reduce these basis-truncation effects, improve the
separation of the low-lying levels and potentially allow stable
eigenvectors to be determined at shorter Euclidean times, where the
statistical signal is more precise.
 \section*{Acknowledgements}
L.~B. received support through the German Research Foundation (DFG) through the research unit FOR 5269 “Future
methods for studying confined gluons in QCD.”
The authors gratefully acknowledge the Gauss Centre for Supercomputing e.V. (www.gauss-centre.eu) for funding this project by providing computing time on the GCS Supercomputer SuperMUC-NG at Leibniz Supercomputing Centre (www.lrz.de). Additional simulations were performed on the QPACE 3 cluster at the University of Regensburg.
 
\small
\bibliographystyle{abbrvnat}
\bibliography{references}

\begin{thebibliography}{15}
\providecommand{\natexlab}[1]{#1}
\providecommand{\url}[1]{\texttt{#1}}
\expandafter\ifx\csname urlstyle\endcsname\relax
  \providecommand{\doi}[1]{doi: #1}\else
  \providecommand{\doi}{doi: \begingroup \urlstyle{rm}\Url}\fi

\bibitem[Alarc{\'o}n(2021)]{Alarcon:2021dlz}
J.~M. Alarc{\'o}n.
\newblock {Brief history of the pion{\textendash}nucleon sigma term}.
\newblock \emph{Eur. Phys. J. ST}, 230\penalty0 (6):\penalty0 1609--1622, 2021.
\newblock \doi{10.1140/epjs/s11734-021-00145-6}.

\bibitem[Alexandrou et~al.(2024)]{Alexandrou:2024tin}
C.~Alexandrou et~al.
\newblock {Investigation of pion-nucleon contributions to nucleon matrix
  elements}.
\newblock \emph{Phys. Rev. D}, 110\penalty0 (9):\penalty0 094514, 2024.
\newblock \doi{10.1103/PhysRevD.110.094514}.

\bibitem[Aoki et~al.(2026)]{FLAG2024}
Y.~Aoki et~al.
\newblock {FLAG review 2024}.
\newblock \emph{Phys. Rev. D}, 113\penalty0 (1):\penalty0 014508, 2026.
\newblock \doi{10.1103/nfzp-p5dn}.

\bibitem[Bali et~al.(2023)]{RQCD:2022xux}
G.~Bali et~al.
\newblock {Scale setting and the light baryon spectrum in N$_{f}$ = 2 + 1 QCD
  with Wilson fermions}.
\newblock \emph{JHEP}, 05:\penalty0 035, 2023.
\newblock \doi{10.1007/JHEP05(2023)035}.

\bibitem[Barca(2025)]{Barca:2025det}
L.~Barca.
\newblock {Current-enhanced excited states in lattice QCD three-point
  functions}.
\newblock \emph{Phys. Rev. D}, 112\penalty0 (9):\penalty0 L091503, 2025.
\newblock \doi{10.1103/69yc-d74z}.

\bibitem[Barca(2026)]{Barca:2026juc}
L.~Barca.
\newblock {Evidence of current-enhanced excited states in lattice QCD
  three-point functions}.
\newblock \emph{PoS}, LATTICE2025:\penalty0 004, 2026.
\newblock \doi{10.22323/1.518.0004}.

\bibitem[Barca et~al.(2023)]{Barca:2022uhi}
L.~Barca et~al.
\newblock {Toward N to N{\ensuremath{\pi}} matrix elements from lattice QCD}.
\newblock \emph{Phys. Rev. D}, 107\penalty0 (5):\penalty0 L051505, 2023.
\newblock \doi{10.1103/PhysRevD.107.L051505}.

\bibitem[Barca et~al.(2025)]{Barca:2024hrl}
L.~Barca et~al.
\newblock {Nucleon sigma terms with a variational analysis from Lattice QCD}.
\newblock \emph{Phys. Rev. D}, 111\penalty0 (3):\penalty0 L031505, 2025.
\newblock \doi{10.1103/PhysRevD.111.L031505}.

\bibitem[Barca et~al.(2026)]{Barca:2025dca}
L.~Barca et~al.
\newblock {Investigating a two-level algorithm for fermionic observables}.
\newblock \emph{Phys. Rev. D}, 113\penalty0 (3):\penalty0 034505, 2026.
\newblock \doi{10.1103/grxd-71sk}.

\bibitem[Briceno et~al.(2017)]{Briceno:2016mjc}
R.~A. Briceno et~al.
\newblock {Isoscalar $\pi\pi$ scattering and the $\sigma$ meson resonance from
  QCD}.
\newblock \emph{Phys. Rev. Lett.}, 118\penalty0 (2):\penalty0 022002, 2017.
\newblock \doi{10.1103/PhysRevLett.118.022002}.

\bibitem[Gupta et~al.(2021)]{Gupta:2021ahb}
R.~Gupta et~al.
\newblock {Pion{\textendash}Nucleon Sigma Term from Lattice QCD}.
\newblock \emph{Phys. Rev. Lett.}, 127\penalty0 (24):\penalty0 242002, 2021.
\newblock \doi{10.1103/PhysRevLett.127.242002}.

\bibitem[Hoferichter et~al.(2015)]{Hoferichter:2015dsa}
M.~Hoferichter et~al.
\newblock {High-Precision Determination of the Pion-Nucleon
  {\ensuremath{\sigma}} Term from Roy-Steiner Equations}.
\newblock \emph{Phys. Rev. Lett.}, 115:\penalty0 092301, 2015.
\newblock \doi{10.1103/PhysRevLett.115.092301}.

\bibitem[Hoferichter et~al.(2023)]{Hoferichter:2023ptl}
M.~Hoferichter et~al.
\newblock {On the role of isospin violation in the pion{\textendash}nucleon
  {\ensuremath{\sigma}}-term}.
\newblock \emph{Phys. Lett. B}, 843:\penalty0 138001, 2023.
\newblock \doi{10.1016/j.physletb.2023.138001}.

\bibitem[Jones et~al.(2026)]{Pia:lattice2026}
P.~L. Jones et~al.
\newblock {Direct Determination of the Sigma Terms of the Baryon Octet from
  $N_f=2+1$ Lattice QCD with Wilson fermions}.
\newblock Work in progress, 2026.

\bibitem[Liang et~al.(2025)]{Liang:2025adz}
Z.-R. Liang et~al.
\newblock {Two-Loop Extraction of the Pion-Nucleon Sigma Term}.
\newblock 2025.
\newblock {arXiv:2508.11435 [hep-ph]}.

\end{thebibliography}

\end{document}